\documentclass[superscriptaddress,aip,amsmath,amssymb,floatfix,reprint,raggedbottom]{revtex4-1}
\usepackage[pdftex]{graphicx}
\usepackage{amsmath}
\usepackage{lipsum}
\usepackage{color}
\usepackage{soul}
\usepackage{balance}
\usepackage{fancyhdr}
\usepackage{booktabs}
\usepackage{enumitem}
\usepackage{listings}
\usepackage{hyperref}

\hypersetup{colorlinks=true,linkcolor=blue,citecolor = blue,urlcolor = blue, linktocpage}

\fancypagestyle{firststyle}
{
\fancyhf{}

\rfoot{\thepage}
\pagestyle{fancy}
}

\usepackage{fancyhdr}

\begin{document}

\title{Benchmarking a Probabilistic Coprocessor}
\author{Jan Kaiser}
\email{kaiser32@purdue.edu}

\affiliation{Elmore Family School of Electrical and Computer Engineering, Purdue University, West Lafayette, IN, 47906 USA}

\author{Risi Jaiswal}
\affiliation{Elmore Family School of Electrical and Computer Engineering, Purdue University, West Lafayette, IN, 47906 USA}

\author{Behtash Behin-Aein}
\affiliation{Ludwig Computing, 516 Northwestern Ave, West Lafayette, IN, 47906 USA}

\author{Supriyo Datta}
\affiliation{Elmore Family School of Electrical and Computer Engineering, Purdue University, West Lafayette, IN, 47906 USA}
\date{\today}

\begin{abstract}
Computation in the past decades has been driven by deterministic computers based on classical deterministic bits. Recently, alternative computing paradigms and domain-based computing like quantum computing and probabilistic computing have gained traction. While quantum computers based on q-bits utilize quantum effects to advance computation, probabilistic computers based on probabilistic (p-)bits are naturally suited to solve problems that require large amount of random numbers utilized in Monte Carlo and Markov Chain Monte Carlo algorithms. These Monte Carlo techniques are used to solve important problems in the fields of optimization, numerical integration or sampling from probability distributions. However, to efficiently implement Monte Carlo algorithms the generation of random numbers is crucial. In this paper, we present and benchmark a probabilistic coprocessor based on p-bits that are naturally suited to solve these problems. We present multiple examples and project that a nanomagnetic implementation of our probabilistic coprocessor can outperform classical CPU and GPU implementations by multiple orders of magnitude. 

\end{abstract}
\pacs{}
\maketitle
\thispagestyle{firststyle}
\section{Introduction}

Due to the slowing down of Moore's law, unconventional computing approaches that move algorithms closer to physics are of great current interest. One promising field is that of probabilistic computing which has achieved wide spread attention \cite{goto_combinatorial_2019,aramon_physics-inspired_2019,yamamoto_73_2020,yamaoka_243_2015,ahmed_probabilistic_2020,patel_ising_2020}. One form of probabilistic computing inspired by Feynman\cite{feynman_simulating_1982} makes use of a compact probabilistic (p-)bit that can fast and efficiently generate random numbers at an area footprint of only 3 transistors and 1 stochastic magnetic tunnel junction (s-MTJ) \cite{camsari_stochastic_2017,camsari_implementing_2017,borders_integer_2019}. In comparison, CMOS alternatives require 1000+ transistors to perform a similar functionality \citep{borders_integer_2019}. These p-bits can be utilized to compactly implement new and existing algorithms following Monte Carlo methods. Monte Carlo methods are easy to understand, easy to implement and allow embarrassingly parallel computation. Markov Chain Monte Carlo algorithms are very powerful and have been named as part of the top 10 algorithms of the 20th century \cite{cipra_best_2000}. They have the property that the random number generator (RNG) is crucial for the overall computation and can be the bottleneck since other mathematically heavy parts of data processing is readily addressed by with today's GPUs, resistive crossbars or by pipelining \cite{nguyen_gpu_2007}.

In this paper, we propose and benchmark a probabilistic coprocessor that is naturally suited to address a wide range of tasks that require a large amount of random numbers. The architecture is designed to make the most use out of parallel RNG units that could be implemented with stochastic MTJ based p-bits on an integrated chip. For each problem, we emulate the probabilistic coprocessor with a Field Programmable Gate Array (FPGA) and project the performance of an integrated \textit{p-computer} using a nanomagentic implementations. We compare the coprocessor with conventional CPU and GPU platforms using state-of-the-art software implementations. We show that the \textit{p-computer} emulation can achieve 1-3 orders of magnitude improvement when compared to CPU and GPU implementations whereas the projected \textit{p-computer} can achieve another 2-3 order of magnitude improvement for each problem when compared to the emulation. 

The paper is structured as follows. First, we introduce the probabilistic coprocessor architecture and show how it could be implemented in an integrated circuit using MTJ-based p-bits. We then demonstrate the advantage of parallelization on the simple problem of estimating $\pi$. Then we demonstrate the \textit{p-computer} on an example of uncertainty classification and Bayesian networks. Consequently, we show that Markov Chain Monte Carlo techniques can be implemented as well and show the performance on the Knapsack optimization problem. 

\section{Architecture of the probabilistic coprocessor}

\begin{figure}[h!]
    \setlength\abovecaptionskip{-0.5\baselineskip}
    \centering
    \includegraphics[width=1\linewidth]{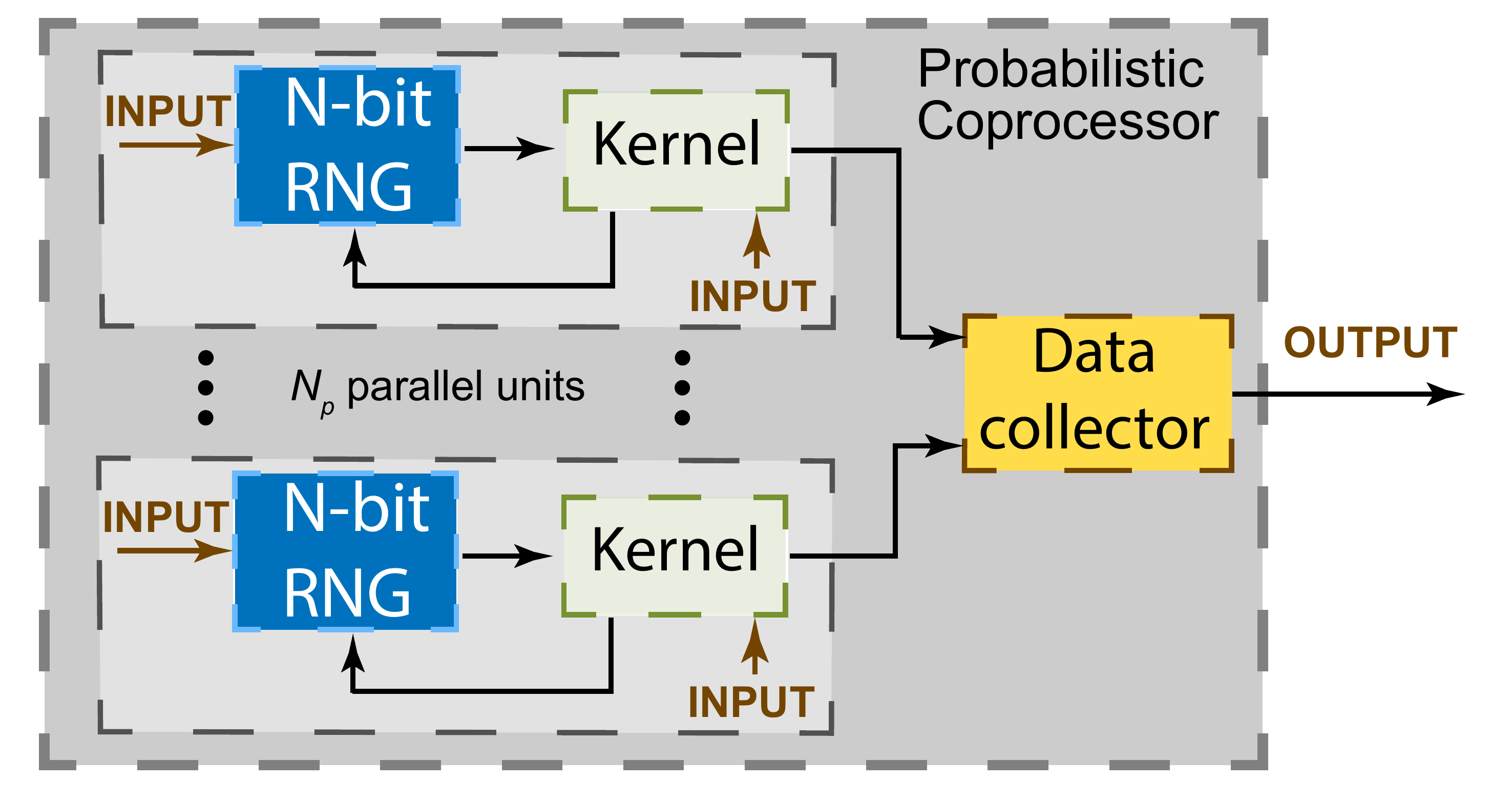} 
    \vspace{0.1in}
    \caption{\textbf{Structure of Probabilistic coprocessor:} \textit{N-bit RNG} and \textit{Kernel} blocks can have parallel units for problems that benefit from it. For \textit{N-bit RNG} blocks, s-MTJ based p-bit arrays can be utilized.}
    \label{fig: ProbCoproc}
\end{figure}

The probabilistic coprocessor is shown in Fig. \ref{fig: ProbCoproc} which describes the general architecture that can solve a wide variety of problems that can be address with probabilistic computing. The random number generator (RNG) block is responsible for fast generation of random numbers which could be implemented in an integrated circuit using the 1 s-MTJ and 3 transistor based p-bits (see section \ref{sec: pbits} for more details). The processing of the generated random numbers is performed by the \textit{Kernel} which will be adapted depending on the problem. The \textit{N-bit RNG} and \textit{Kernel} block is copied in order to run multiple operations in parallel. The data processing is pipelined as much as possible in order to make use of fast RNGs. The processed data is collected in the yellow data collector block. We note that for usual computational pipelines the input is dependent on other processes. Here, since the input into the pipeline is the RNG, we can make full use of pipelining without waiting for the data processing of other processes to finish. In the methods section, the \textit{p-computer} emulation is described in detail.

\subsection{Hardware implementation for \textit{N-bit RNG} block}
\label{sec: pbits}
As hardware implementation for the \textit{N-bit RNG} block in the architecture in Fig. \ref{fig: ProbCoproc}, an array of compact hardware p-bits \cite{camsari_implementing_2017} consisting of just 3 transistors and 1 stochastic MTJ can be utilized whereas CMOS alternatives like linear-feedback shift registers (LFSRs) would require 1000+ transistors. We note that the p-bits can supply tunable random numbers whereas simple LFSRs would require additional hardware in the form of lookup tables (LUTs) to generate tunable random numbers \cite{sutton_autonomous_2020}. The p-bit design was first proposed by Camsari et al. \cite{camsari_implementing_2017} and analyzed by Hassan et al. \cite{hassan_low-barrier_2019,hassan_quantitative_2021}. The overall power consumption of one hardware p-bit is around $20 \mu$W which is orders of magnitude better than CMOS alternatives\cite{borders_integer_2019}. Another advantage of this compact RNG-source is that it is based on stochastic MTJs which can operate at nanosecond speeds which has been shown theoretically as well as experimentally\cite{kaiser_subnanosecond_2019,safranski_demonstration_2021,hayakawa_nanosecond_2021,kanai_theory_2021}. The random number generation of the p-bit design can be tuned with the input voltage at the gate of the NMOS-transistor $V_{IN}$ and has a bipolar output voltage $V_{OUT}=\pm V_{DD}/2$. The average of the binary bitstream of the p-bit can be described by $\langle V_{OUT} \rangle \approx V_{DD}/2 \cdot \tanh(V_{IN}/V_0)$ where $V_0$ is a material dependent reference voltage. This tunability is important for problems with feedback (compare section \ref{sec: Knapsack}) since the proposal distribution is dependent on the current configuration of the system. For cases where the proposed configuration must be close to the current p-bit state vector, all p-bits can be biased to the current configuration and only have a small probability to generate a perturbed state.

\begin{figure}[h!]
    \setlength\abovecaptionskip{-0.5\baselineskip}
    \centering
    \includegraphics[width=0.7\linewidth]{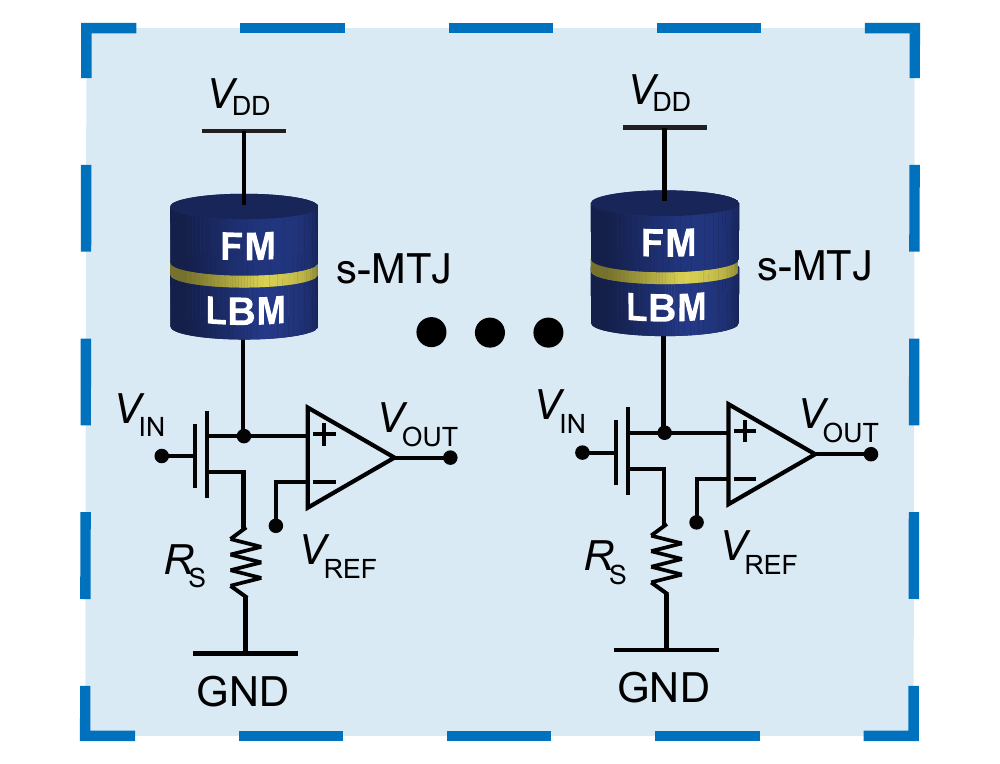} 
    \vspace{0.1in}
    \caption{\textbf{MTJ-based p-bits:} 3T/1MTJ based p-bit that can be used to implemented the \textit{N-bit RNG} block compactly and efficiently.}
\end{figure}

\section{Benchmarking the probabilistic coprocessor}
In this section, we address several problems with the probabilistic coprocessor. We compare the time to solution of our \textit{p-computer} emulation and \textit{p-computer} projection with the performance on CPU and GPU for every problem.

Broadly, there are two classes of Monte Carlo algorithms: simple Monte Carlo algorithms that perform direct sampling and Markov Chain Monte Carlo techniques. Both require many random numbers, however, Markov Chain Monte Carlo also requires feedback from \textit{Kernel} back to \textit{N-bit RNG} (see black arrow in Fig. \ref{fig: ProbCoproc}).

\subsection{Basic Monte Carlo techniques}
The advantage of basic Monte Carlo techniques is that they are easy to parallelize and do not require convergence. The standard error is given by $\epsilon_r=\sigma/\sqrt{N_S}$ and the solution accuracy improves proportional to $1/\sqrt{N_S}$. As mentioned in the introduction the RNG generation can be the bottleneck since the mathematically heavy part is readily addressed by GPUs, resistive crossbars or by pipelining \cite{kogge_architecture_1981}. Hence, the very rapid random number generation of \textit{p-computers} can be fully utilized. In our \textit{p-computer} emulation, we use pipelining to increase the data processing rate but other approaches are valid.

\subsubsection{Monte Carlo integration}
\begin{figure}[h!]
    \setlength\abovecaptionskip{-0.5\baselineskip}
    \centering
    \includegraphics[width=1\linewidth]{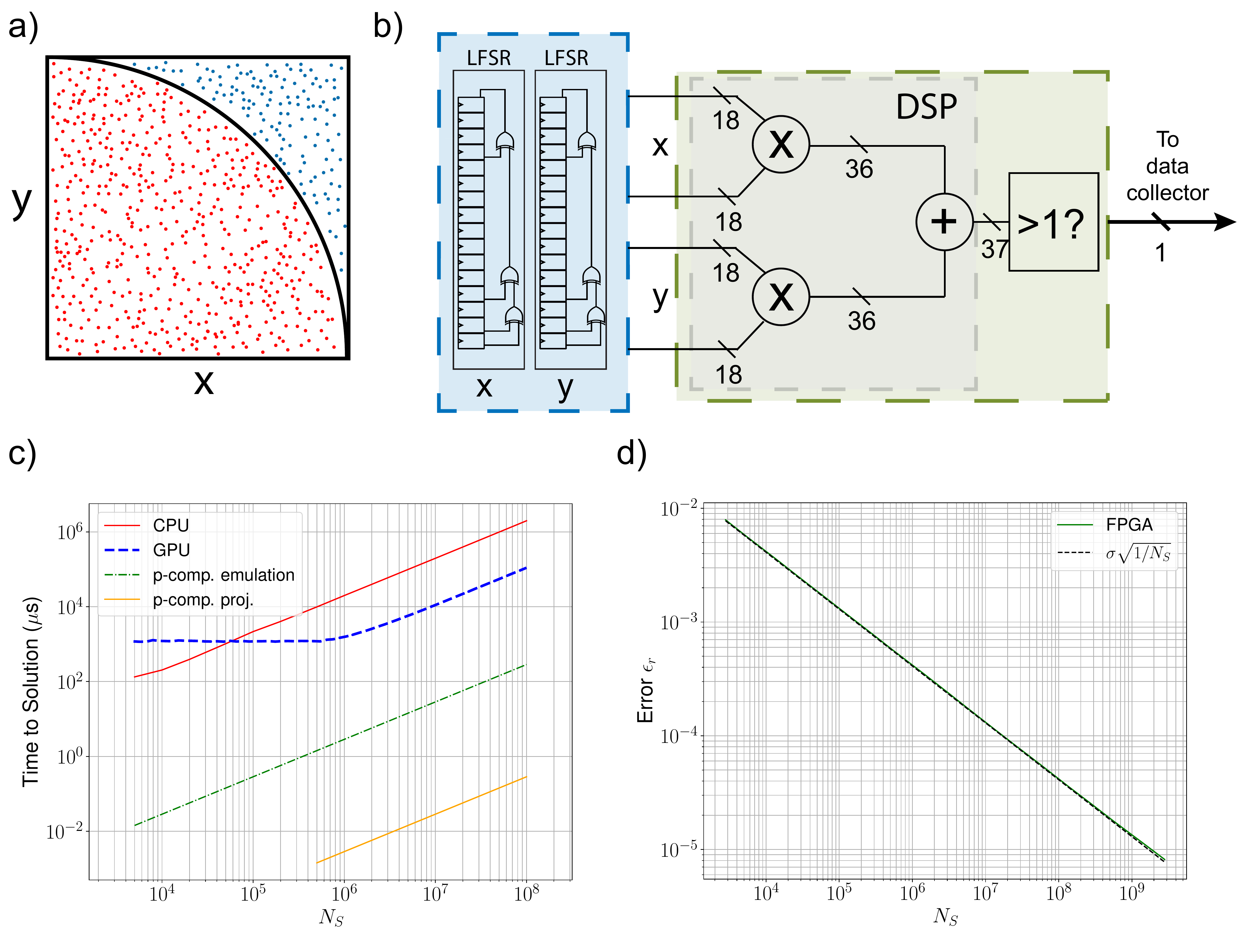} 
    \vspace{0.1in}
    \caption{\textbf{Monte Carlo Calculation of $\pi$:} a) Illustration of the sampling process. Red dots are counted inside the circle, blue dots outside. b) \textit{p-computer} emulation of one $\pi$ sampling block. Linear-feedback shift registers (LFSRs) are used to generate random numbers. The following logic is used to calculate the radius and test if the generate coordinate is inside the circle. c) Time to solution vs. number of samples for CPU, GPU, \textit{p-computer} emulation and projected \textit{p-computer}. d) Accuracy of $\pi$ calculation vs. Number of samples. The result is independent on the platform. Here, \textit{p-computer} emulation results are plotted for different $N_S$. Each point is averaged over $10^5$ samples.}
    \label{fig: Pi}
\end{figure}

One main advantage of the Monte Carlo calculation is that it can be straight-forwardly parallelized. Monte Carlo integration is especially advantageous to use for problems with many dimensions where it scales better than deterministic numerical integration techniques \cite{joseph_markov_2020}. To illustrate this point we demonstrate numerical integration on a simple example of calculating $\pi$ by randomly drawing coordinates from a square as shown in Fig. \ref{fig: Pi} a). For every sample we check if the sample is inside the circle. We then get an approximation of $\pi \approx \frac{4 N_{in}}{N_{all}}$ where $N_{in}$ is the number of points inside the circle and $N_{all}$ is the number of all samples that are collected. The procedure requires to generate many random numbers in order to get as many random coordinate points as possible. In our emulation of the probabilistic coprocessor 2800 \textit{N-bit RNG} /\textit{Kernel} blocks (Fig. \ref{fig: Pi} b)) are implemented in parallel. Here, for each dimension a 18-bit random number between 0 and 1 is generated. For the basic Monte Carlo techniques we plot the time to solution (TTS) dependent on the sample number $N_S$ in Fig. \ref{fig: Pi} c). The more accurate the solution should be the more samples $N_S$ are needed. For the \textit{p-computer} emulation, the TTS plot can be understood as follows: every clock cycle 2800 samples are generated. This results in a linear relationship between numbers of samples and the time to solution:
\begin{equation}
\mathrm{TTS}=\frac{N_S}{f_{clk} \cdot N_{p}}
\label{eq: TTS}
\end{equation}
where $N_S$ is the number of samples, $f_{clk}$ the clock frequency and $N_{p}=2800$ the number of parallel blocks. Eq. \ref{eq: TTS} is general for all Basic Monte Carlo in this section. The accuracy of the $\pi$ estimation is shown in Fig. \ref{fig: Pi} d). For the $\pi$ problem the standard deviation $\sigma$ of the standard error is given by $\sigma=\sqrt{\pi(1-\pi/4)/4}$. The main advantage of \textit{p-computer} compared to CPU and GPU is that the number of parallel blocks $N_{p}$ is much higher. We note, however, that in a fully pipelined design, some setup time $t_{su}$ is required to fill the pipeline. However, in this paper the pipeline stage number is small $t_{su}\ll \mathrm{TTS}$ and $t_{su}$ can be neglected.

We note that this is just a simple example of Monte Carlo integration and other deterministic approaches to calculate $\pi$ are superior. However, for integrals with many dimensions, Monte Carlo integration has better scaling than conventional methods such as trapezoidal and Simpson's rule \cite{joseph_markov_2020}.  

\subsubsection{Uncertainty quantification}
\begin{figure}[hb!]
    \setlength\abovecaptionskip{-0.5\baselineskip}
    \centering
    \includegraphics[width=1\linewidth]{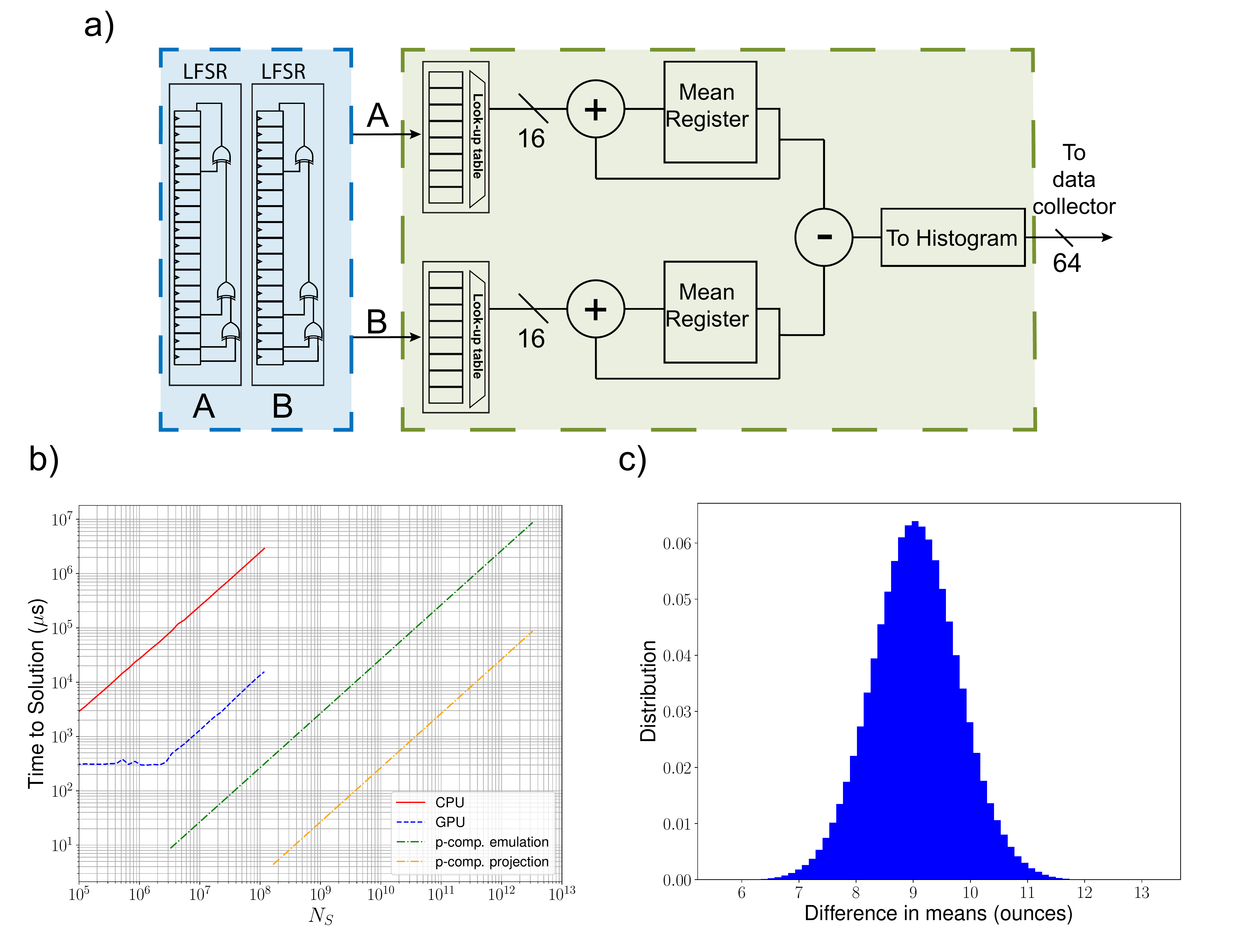} 
    \vspace{0.1in}
    \caption{\textbf{Bootstrap resampling:} a) \textit{p-computer} emulation on sampling block of \textit{p-computer}: LFSRs generate random numbers and the following logic is used to calculate the mean of a given Bootstrap sample, subtract both groups and collect the data in a histogram. b) Time to solution vs. Number of samples $N_S$ for the different platforms. c) Extracted histogram from the \textit{p-computer} emulation using 64 bins and $1.6 \cdot 10^{12}$ samples}
    \label{fig: Bootstrap}
\end{figure}

Fig. \ref{fig: Bootstrap} shows uncertainty quantification using our probabilistic coprocessor. One way of quantifying uncertainty is to use Bootstrap resampling \cite{efron_introduction_1994,efron_second_2003}. Given a limited dataset, this resampling technique can be used to make a statement about the sampling error of dataset \textit{without making assumptions about the underlying distribution}. The Bootstrap technique redraws from the limited dataset with replacement and checks if the parameters like mean or standard deviation change significantly for each resampled distribution. We note note that the ideal Bootstrapping becomes intractable quickly due to the large number of possible redraws which makes Monte Carlo techniques necessary.

Here, we use a dataset\footnote{https://github.com/datascienceforall/dsfa-2018sp-public/tree/master/textbook} for A/B testing containing 1174 mothers and the birth weight of their babies that are classified into two groups: smoker and non-smokers. To quantify if the birth weight is influenced by the fact that the mothers smoked during pregnancy, bootstrap resampling can be utilized. The overall implemented structure on for the \textit{p-computer} emulation is shown in Fig. \ref{fig: Bootstrap} a) where the \textit{Kernel} is adapted to collect the histogram for the difference of the means. Here, instead of reading one number, 64 values for 64 bins are read out. In order to be able to collect statistics for multiple datasets the bin width and bin position for the bootstrapped distribution can be sent to the \textit{p-computer} after implementation of the design. The TTS of the Bootstrap resampling is plotted for different total number of samples and compared to CPU and GPU. In addition, the projection of a \textit{p-computer} with increased number of parallel blocks is plotted in Fig \ref{fig: Bootstrap} b). It can be clearly seen that the \textit{p-computer} emulation performs much better than CPU and GPU implementation. The overall behavior can be described by Eq. \ref{eq: TTS}. Note that the parallelization in this example is done by having $N_{p}=1500$ blocks shown in Fig. \ref{fig: Bootstrap} a) and the mean is calculated serially. In principle, it is possible to have parallelization in both mean calculations which would increase the advantage when comparing the execution to CPU and GPU.

The difference of the means is plotted in a histogram \ref{fig: Bootstrap} c). Using the generated distribution the uncertainty interval of the difference of means between the two groups can be quantified. 

\subsubsection{Bayesian networks}
\begin{figure}[h!]
    \setlength\abovecaptionskip{-0.5\baselineskip}
    \centering
    \includegraphics[width=1\linewidth]{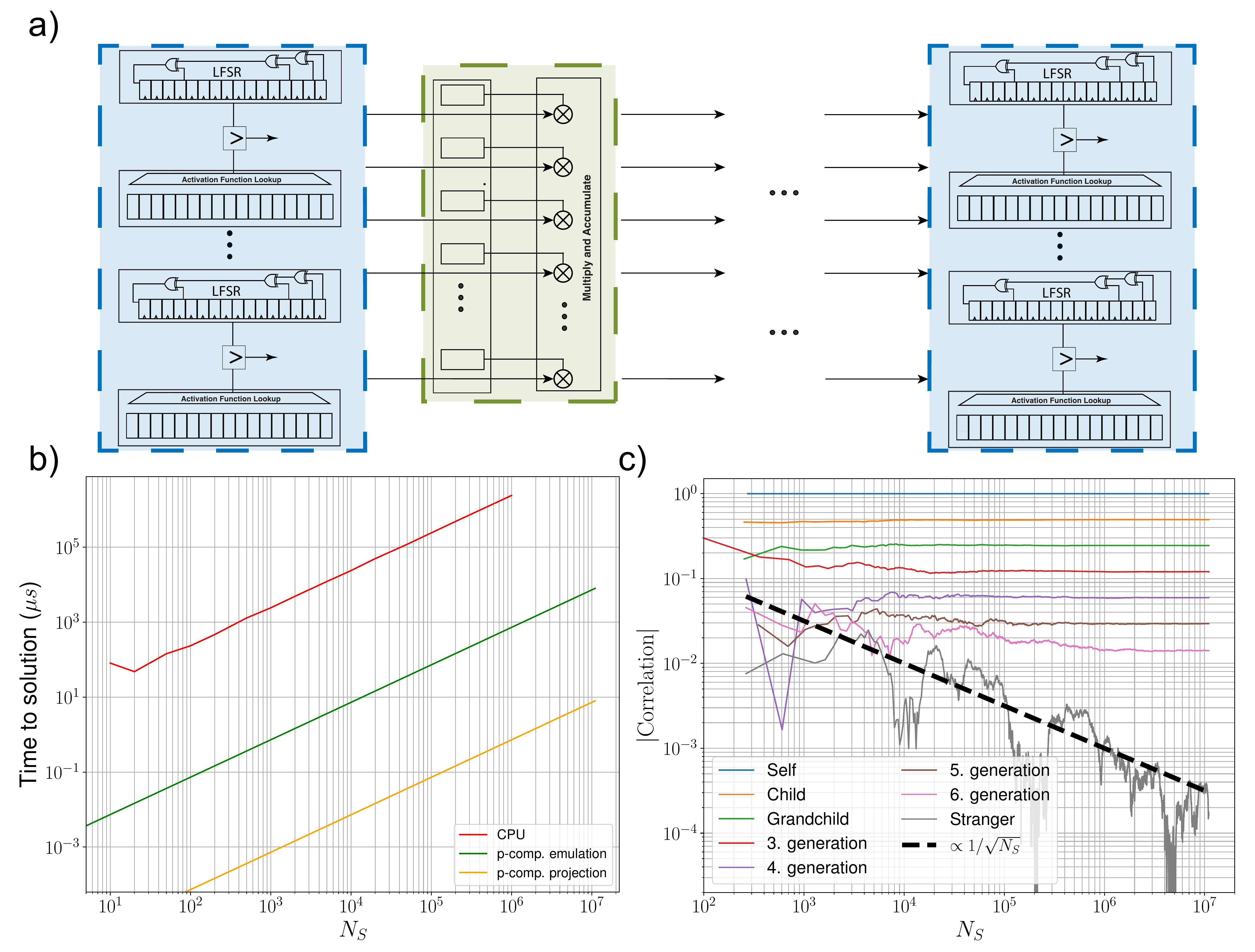} 
    \vspace{0.1in}
    \caption{\textbf{Bayesian network with probabilistic coprocessor:} a) p-computer emulation architecture of Bayesian network. \textit{N-bit RNG} and \textit{Kernel} blocks are routed in series. b) TTS vs $N_S$ for different platforms. c) Absolute value of correlation for a family tree with 7 generations.}
    \label{fig: Bayes}

\end{figure}

Bayesian networks can be mapped to the \textit{p-computer} as shown in Fig. \ref{fig: Bayes} a). Here, multiple \textit{N-bit RNG} and \textit{Kernel} are put in series to represent different layers of a Bayesian network. Every node of the Bayesian network can be represented by a tunable 1-bit RNG or p-bit\cite{faria_implementing_2018}. As an example we use the genetic relatedness in a family tree\cite{faria_implementing_2018,camsari_p-bits_2019,faria_hardware_2021}. The relatedness of two family members can directly be extracted from the network by measuring the correlation of the two corresponding p-bits. The absolute values of the correlation are shown in Fig. \ref{fig: Bayes} b) for a family tree of 7 generations starting with 64 nodes in the first layer down to 1 node in layer 7. At total 127 members of the family tree are represented in the p-bit network. Every network is copied 10 times so that we emulated a network of 1270 p-bits total. Again the error of the correlation goes down $\sim 1/\sqrt{N_S}$ using this method.  Compared to deterministic algorithms, the advantage of a probabilistic computer is that during integration,
 \begin{equation}
P_A (x_A) = \int dx_B P(x_A,x_B),
\label{eqn: eqn3}
\end{equation}
observed variables $x_A$ can directly be measured without integrating over unobserved variables $x_B$ as noted by Feynman\cite{feynman_simulating_1982}.

The Bayesian network implemented on the emulated \textit{p-computer} is benchmarked against the same algorithm run on CPU in Fig. \ref{fig: Bayes} c). Compared to CPU we see around 2 orders of magnitude improvement and expect further improvement for the projected \textit{p-computer} utilizing s-MTJs.

\subsection{Markov Chain Monte Carlo techniques}
For Markov Chain Monte Carlo (MCMC) techniques like the Metropolis-Hastings algorithm, the data processing plays a more important role since a feedback loop is added between the \textit{N-bit RNG} and the \textit{Kernel} block \cite{metropolis_equation_1953,hastings_monte_1970}. This means that parallelism is less effective. However, Markov Chain Monte Carlo algorithms can be more powerful in tasks like optimization and sampling especially in high dimensional spaces. For MCMC other parallelization techniques can be utilized to increase performance like partitioning the random variables into independent groups, multiple-try Metropolis \cite{byrd_reducing_2008,liu_multiple-try_2000} or parallel tempering\cite{swendsen_replica_1986-1,earl_parallel_2005}. 

An important class of problems that can be readily addressed are optimization problems like finding the ground state of an Ising Model or the Knapsack problem. These problems are NP-hard which means that it takes exponential amount of resources to compute a solution. However, Monte Carlo algorithms utilizing random numbers can find approximate solution faster than the deterministic counterpart. To show that our probabilistic coprocessor can be used to solve problems like these, we use it to solve the Knapsack problem. 

\subsubsection{0-1 Knapsack problem}
\label{sec: Knapsack}
\begin{figure}[h!]
    \setlength\abovecaptionskip{-0.5\baselineskip}
    \centering
    \includegraphics[width=1\linewidth]{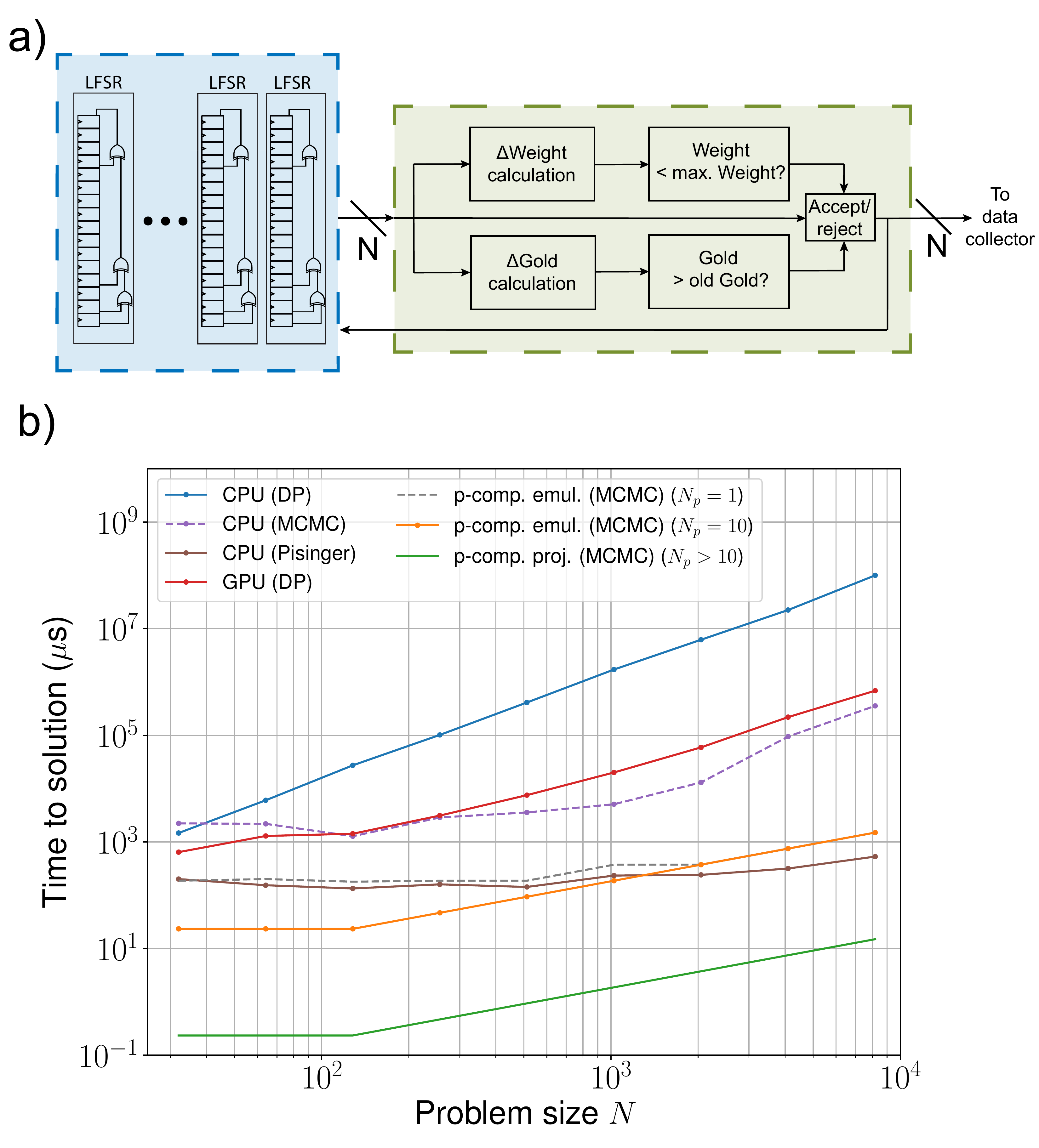} 
    \vspace{0.1in}
    \caption{\textbf{Knapsack problem:} a) \textit{p-computer} emulation on sampling block of \textit{p-computer} : LFSRs generate random numbers and the following logic is used to calculate the mean of a given Bootstrap sample, subtract both groups and collect the data in a histogram. b) Time to solution vs. problem size $N$ for different platforms (CPU, GPU and \textit{p-computer} ) and different algorithms (DP, MCMC, Pisinger \cite{martello_dynamic_1999}).}
    \label{fig: Knapsack}
\end{figure}

The Knapsack problem is a famous difficult optimization problem. It can be mapped to problems like financial modeling, inventory management, resource allocation or yield management \cite{pisinger_minimal_1995}. Given $N$ items where each item has a given weight $W$ and value $V$, the goal is to find a configuration $x = \textrm{max} [\sum_i V_i x_i]$ where $\sum_i W_i x_i \leq C$ with capacity $C$. To utilize the parallel RNG to the fullest we have implemented a pipeline that accepts new proposals every clock cycle as shown in Fig. \ref{fig: Knapsack} a) even if the \textit{Kernel} needs more than one clock cycle to propagate a new proposal. This approach is similar to multiple-try Metropolis \cite{byrd_reducing_2008,liu_multiple-try_2000}. Each proposal uses the current state at a starting point and changes the state of 2 items at random. The generated proposal is then send into the pipeline and the new weight and value is calculated by making use of the fact that only the 2 flipped item's weight and value are changed. If the weight value is smaller or equal to the maximum allowed weight, the new state is propagated to the accept/reject block where it is weighted by an exponential function and compared to the last accepted state. If accepted the current state is updated.

In Fig. \ref{fig: Knapsack} b) the TTS of the Knapsack problem for different items numbers $N$ is compared to the CPU and GPU implementations utilizing different algorithms like MCMC or dynamic programming (DP). TTS for this optimization problem for the stochastic MCMC approach is defined by the commonly used formula \cite{ronnow_defining_2014-1,mcmahon_fully_2016} with $99 \%$ success probability,
\begin{equation}
TTS(N)=T_{run} \cdot \frac{\log(0.01)}{\log(1-p_s)},
\end{equation}
where $p_s$ is the probability of finding an acceptable solution when running the algorithm for a time $T_{run}$. We define a success as finding a configuration that has a value that is larger than $99 \%$ of the ideal solution value. The emulated p-computer uses 10 different Markov Chains in parallel which decreases the TTS for smaller problem sizes (compare orange solid line to grey dashed line). The running time $T_{run}$ is given by $T_{run}=N_S/f_{clk}$ for the \textit{p-computer} emulation and as CPU running time for CPU. For the MCMC scheme we use a simulated annealing approach where the temperature is reduced exponentially by a factor of two every $N_S/10$ samples. To find the best annealing schedule for every problem size $N$, we varied the number of total samples $N_S=2^x \cdot 10$ with $x=1,2,...20$ and and plotted the minimum TTS in Fig. \ref{fig: Knapsack} b). For each item property, we randomly draw a value $V_i$ and weight $W_i$ between 0 and 1000. As capacity we use $C=\sum_{i=0}^N W_i/2$ for all problems sizes.

Fig. \ref{fig: Knapsack} b) shows that \textit{p-computer} outperforms the dynamic programming algorithm on both CPU and GPU and is also about 1-2 orders faster than the MCMC implementation on CPU. The \textit{p-computer} emulation is, however, outperformed by the special deterministic \textit{combo} algorithm developed by Pisinger and colleagues \cite{martello_dynamic_1999,kellerer_knapsack_2004}. We note that the MCMC approach is much more general than this specific algorithm. In addition, the algorithm on the probabilistic computer could be further improved for example by adding parallel tempering\cite{swendsen_replica_1986-1,earl_parallel_2005}.

\section{Conclusion}
In conclusion, we have benchmarked a probabilistic coprocessor emulated on FPGA for multiple Monte Carlo applications. We show that the coprocessor can outperform conventional platforms by multiple orders of magnitude. We project that an integrated chip based on stochastic MTJs could show even better improvement in power consumption and speed.

\acknowledgements
This work was supported in part by ASCENT, one of six centers in JUMP, a Semiconductor Research Corporation (SRC) program sponsored by DARPA.

\bibliographystyle{naturemag}
\balance\bibliography{library}

\section{Methods}
For CPU benchmarks we used a Intel(R) Xeon(R) @ 2.3GHz CPU using C++ implementations. GPU results were obtained on a Tesla T4 @ 1.59GHz GPU using CUDA or Tensorflow platforms. For CPU, we note that interfaces like \textit{OpenMP} can be utilized to run the Monte Carlo approach on multiple CPU cores. Here, we did not utilize any multi-processor interfaces for the CPU benchmarks. In general, the number of FPGA integration cores can be much larger than the typical number of CPU or GPU cores.

\subsection{Emulated \textit{p-computer} implementation}
\label{sec: FPGA}

\begin{figure}[h!]
    \setlength\abovecaptionskip{-0.5\baselineskip}
    \centering
    \includegraphics[width=1\linewidth]{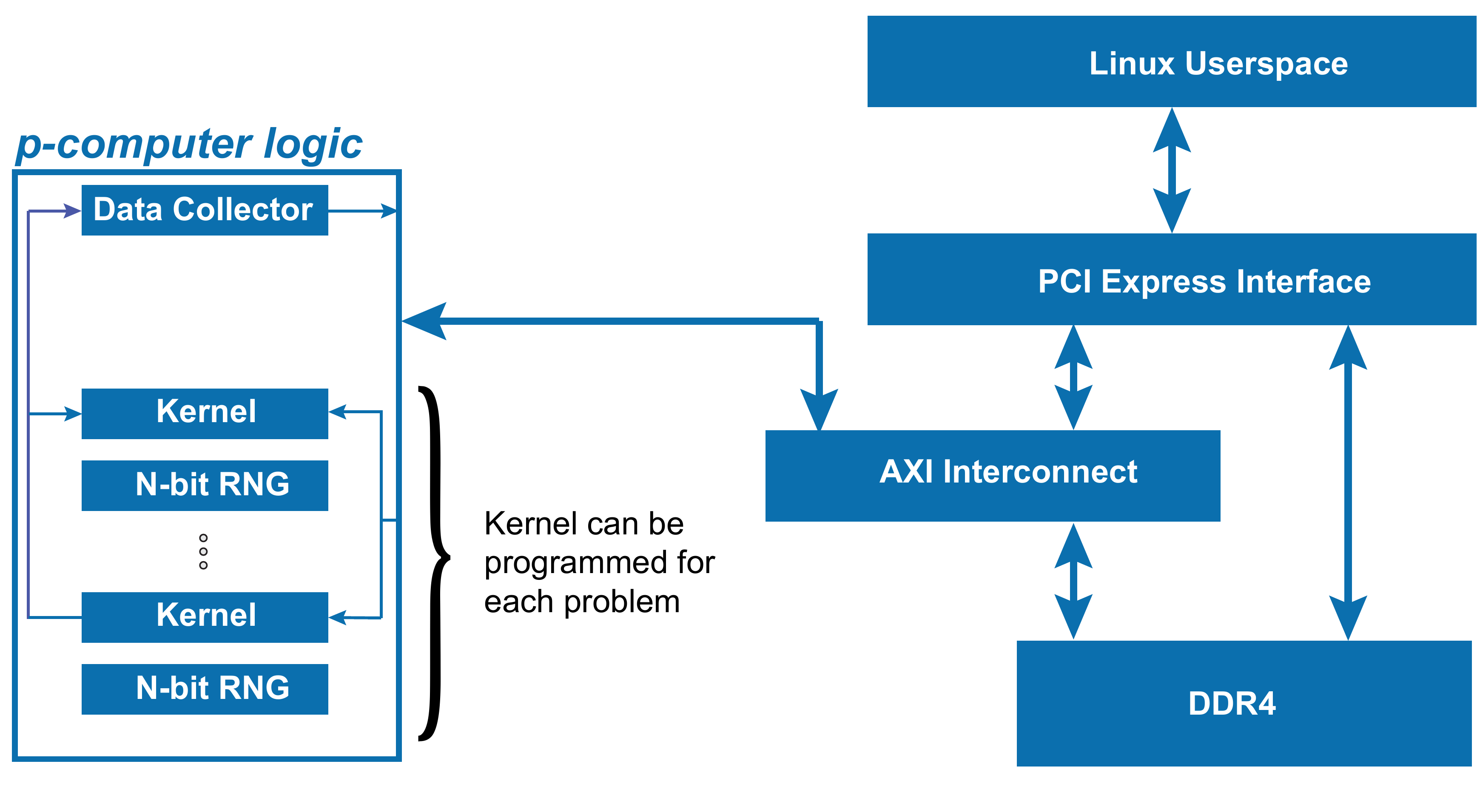} 
    \vspace{0.1in}
    \caption{\textbf{Logical Organization and Memory Map of the FPGA used to emulate the \textit{p-computer} :} The \textit{p-computer} is connected to the AWS userspace through PCI Express interface (PCIe). The data generated by the \textit{p-computer} is saved on the DDR4 on-chip memory.}
    \label{fig: FPGA}
\end{figure}

As FPGA hardware, a Xilinx Virtex Ultrascale+ \textit{xcvu9p-flgb2104-2-i} provided via Amazon Web Services F1 cloud-accessible EC2 compute instances is utilized. The logical organization of the FPGA is shown in Fig. \ref{fig: FPGA}. The problem parameters can be sent from the Linux userspace to the \textit{p-computer} through the PCIe interface. For control signals, the AXI-Lite bus for direct register access is utilized. The results from the \textit{p-computer} operation shown in Fig. \ref{fig: ProbCoproc} is sent to the on-chip DDR4 memory using an AXI interconnect interface. The data can then be read out from the DDR4 and send to the Linux Userspace through PCIe using direct memory access (DMA). For each data point the time stamp is saved in order to be able to extract the execution time of the emulated \textit{p-computer} accurately.

\end{document}